# Equation of States for Elections

Bih-Yaw Jin

Department of Chemistry and Center of Theoretical Sciences, National Taiwan University, Taipei, Taiwan 10617, R.O.C.

In the US 2008 presidential election, Barack Obama was elected as the 44th president of United States, winning the 53% of popular votes and 68% of electoral votes; in the election of 2000, Al Gore lost the election receiving 49 % of electoral votes, although he had more popular votes. It is generally believed that the electoral votes and the popular votes are correlated; however the detailed quantitative relationship for these two quantities is unclear. Here, we found an interesting relationship between fractions of electoral votes and fractions of popular votes in the presidential elections of the United States by examining the election results from 1932 to 2004. Moreover, this curve could provide an interesting explanation for the results of other elections that have taken place in Taiwan.

To give a quantitative analysis for the results of US presidential elections, we introduce two state variables, the fraction of popular votes for republican, x, i.e. the ratio of popular votes obtained by the republican's presidential candidate and total popular votes. The fraction of electoral votes for republican, y is similarly defined. Note that x and y are very similar to the concepts of mole fractions in chemistry. In figure 1, we plot the results of US presidential election from 1932 to 2004 (blue circles). It is apparent that the distribution of these data points has a strong resemblance to the titration curve commonly used in the acid-base equilibrium as described by the Henderson-Hasselbalch equation, i.e. pH=pKa+log([base]/[acid]) [1]. Thus, we propose that the generalized Henderson-Hasselbalch equation, $\xi(x-0.5)=\log[y/(1-y)]$, can be used as the equation of states for any particular electoral system. The fitting parameter, $\xi$, for the US election is found to be 16.12. Interestingly, the result of the 2008 election falls beautifully on the same curve (blue diamond). However, there are fluctuations around the fitting curve, explaining why G. W. Bush was elected, while Al Gore had more popular votes.

Incidentally, at the beginning of 2008, in the re-election of representatives for the parliament in Taiwan, the-then opposition party, Kuomintang, had a landslide victory by winning about 80% of the district seats with only 58% of the total votes based on a newly adopted electoral system, in which only one representative get elected for each electoral district. The apparent disproportion raises some puzzles on the fairness of the new electoral system. The new electoral system in Taiwan has some similarity to the US presidential election, i.e. the winner get the only seat or all electoral votes of that district or state. Thus plotting the result on the same (x, y) space (red empty square), it is reasonable to see that the distribution of seats in this election falls around the equation of states of the US presidential election.

Additionally, the results of the Taiwan's previous three local elections for governors of counties and townships also fall on the same curve as shown in figure 1 (red solid circles). In the same figure, we have also plotted the results of four previous parliamentary elections (red empty circles) held under the previous electoral system, in which many representatives are elected in each electoral district and each party might get some seats in proportion to the votes each candidate received. These data are not distributed around the previous US curve. But, interestingly, we can fit these data by the generalized Henderson-Hasselbalch equation with a different parameter, $\xi$=2.65. The resulting equation of states has a smaller slope. Thus, we believe that the generalized Henderson-Hasselbalch equation with different value of fitting parameter, $\xi$, can be used to describe the correlation between fraction of total votes and fraction of electoral votes for different electoral systems, when only two major competing parties are involved.

Correspondence should be addressed to Bih-Yaw Jin (e-mail: byjin@ntu.edu.tw).

Figure 1: x-axis: the fraction of popular votes republican's candidate obtained; y-axis: the fraction of electoral votes republican's candidate obtained. Blue circles: US presidential elections from 1932 to 2004; big blue diamond: 2008 US presidential election; red square: 2008 Taiwan representative election of parliament based on the new electoral system; red empty circles: previous three Taiwan representative elections of parliament based on the old electoral system; red solid circle: results of three recent elections for the local county and township governors in Taiwan.

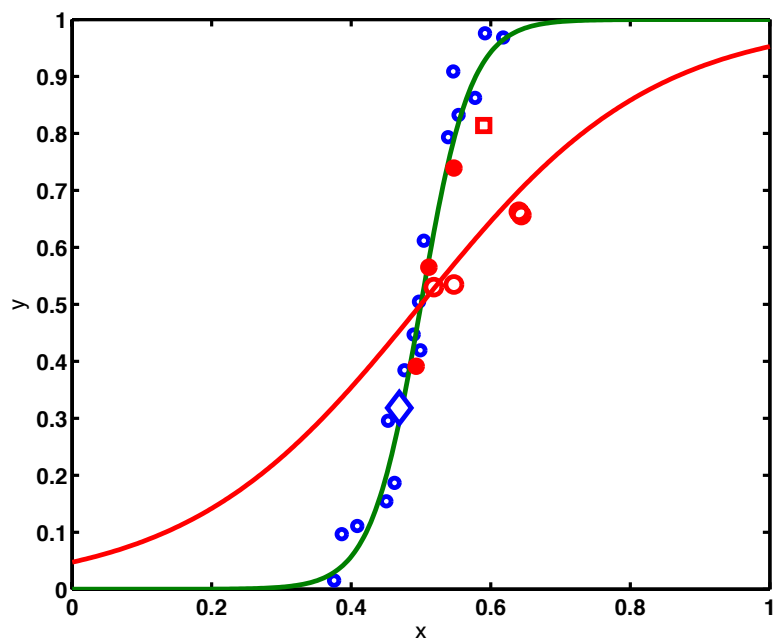